\documentclass[%
 reprint,
 amsmath,amssymb,
 prb,
]{revtex4-2}

\usepackage{graphicx}
\usepackage{dcolumn}
\usepackage{bm}
\usepackage[version=4]{mhchem}
\usepackage{here}
\usepackage{siunitx}

\usepackage{xcolor}

\definecolor{citecolor}{HTML}{191970}
\definecolor{urlcustomcolor}{HTML}{C71585}

\usepackage[
  colorlinks=true,
  linkcolor=black,
  citecolor=citecolor,
  urlcolor=urlcustomcolor
]{hyperref}

\begin{document}

\preprint{APS/123-QED}

\title{Electrical detection of spin-flip transition in metal/\ce{Na_5Co_{15.5}Te_6O_{36}} heterostructure}

\author{Hirotsugu Tagami}
 \thanks{%
  Email: \href{mailto:hirotsugu@g.ecc.u-tokyo.ac.jp}{\textcolor{black}{hirotsugu@g.ecc.u-tokyo.ac.jp}}%
}
\author{Takuya Kawada}
\author{Yuki Shiomi}
\affiliation{%
 Department of Basic Science, The University of Tokyo, Tokyo 153-8902, Japan\\
}%

\date{\today}

\begin{abstract}
We report on the longitudinal magnetoresistance (MR) in thin metal films on an Ising-type antiferromagnetic insulator, \ce{Na_5Co_{15.5}Te_6O_{36}} (\ce{NCTO}). Steep changes in the MR spectra with hysteresis were observed at spin-flip transitions driven by magnetic fields applied along the easy axis of the \ce{NCTO} crystal. The MR jumps almost follow step-like changes in magnetization at the spin-flip transition. At very low temperatures where \ce{Co} moments are partially frozen, the MR anomalies exhibit a tunnel-magnetoresistance-like shape. The observed MR anomalies at the spin-flip transition are attributed to strain effects via magnetostriction upon the magnetic-structure change of the \ce{Co} nets in \ce{NCTO}, because similar MR jumps are observed in \ce{Pt/NCTO}, \ce{Pt/SiO_x/NCTO}, and \ce{Cu/NCTO} heterostructures. Interestingly, we found that the high-field slopes of the MR spectra show opposite signs between \ce{Pt/NCTO} and \ce{Cu/NCTO} at low temperatures. Because similar MR spectra are observed for \ce{Pt/SiO_x/NCTO} in which the interface magnetic interaction is negligible, the weak antilocalization due to the strong spin-orbit interaction of the Pt films is likely to contribute to the low-temperature MR.
\end{abstract}

\maketitle


\section{\label{sec:level1} Introduction}

Heavy metal/magnetic insulator bilayers have attracted considerable attention recently because the magnetic information of the underlying magnetic layer can be detected electrically via the magnetoresistance (MR) of metal layers. A prominent example is the spin Hall magnetoresistance (SMR) effect, which arises from the interaction between the spin polarization accumulated at the interface due to the spin Hall effect in a heavy metal and the magnetization of the magnetic insulator~\cite{Chen2013-bx, Leiviska2025-sz}. The SMR was first discovered in Pt/garnet ferrite bilayers \cite{Nakayama2013-vq,Althammer2013-pi}, but was soon applied to other magnetic materials, including antiferromagnets without net magnetization in a zero magnetic field \cite{Fischer2020-rh,Sugi2023-me,Sando2024-ly}. The interface spin accumulation caused by the spin Hall effect also results in the Hanle MR, which arises from the precession of the accumulated spins by magnetic fields \cite{PhysRevLett.99.126601}, leading to a similar angular dependence of the SMR \cite{Velez2016-er}. Heavy metals with large spin-orbit interaction (SOI), such as \ce{Pt} and \ce{W}, are essential for observing the MR associated with interface spin accumulation.
\par

The target material of this study is an antiferromagnetic insulator, \ce{Na_5Co_{15.5}Te_6O_{36} (NCTO)}. \ce{NCTO} is regarded as a model material for an Ising antiferromagnet \cite{Zhao2022-hh}. This material has a strong Ising anisotropy, leading to a spin-flip transition under magnetic fields applied along its easy axis. To the best of our knowledge, MR studies on metal films deposited on antiferromagnetic insulators with spin-flip transitions are rare, whereas spin-flop transitions have been frequently studied for various antiferromagnets, such as \ce{MnF2} \cite{Wu2016-ib}, \ce{Cr2O3} \cite{Liao2025-hu}, and \ce{NiO} \cite{PhysRevLett.118.147202}. 
\par

\begin{figure*}[th]
    \centering
    \includegraphics[width=1\linewidth]{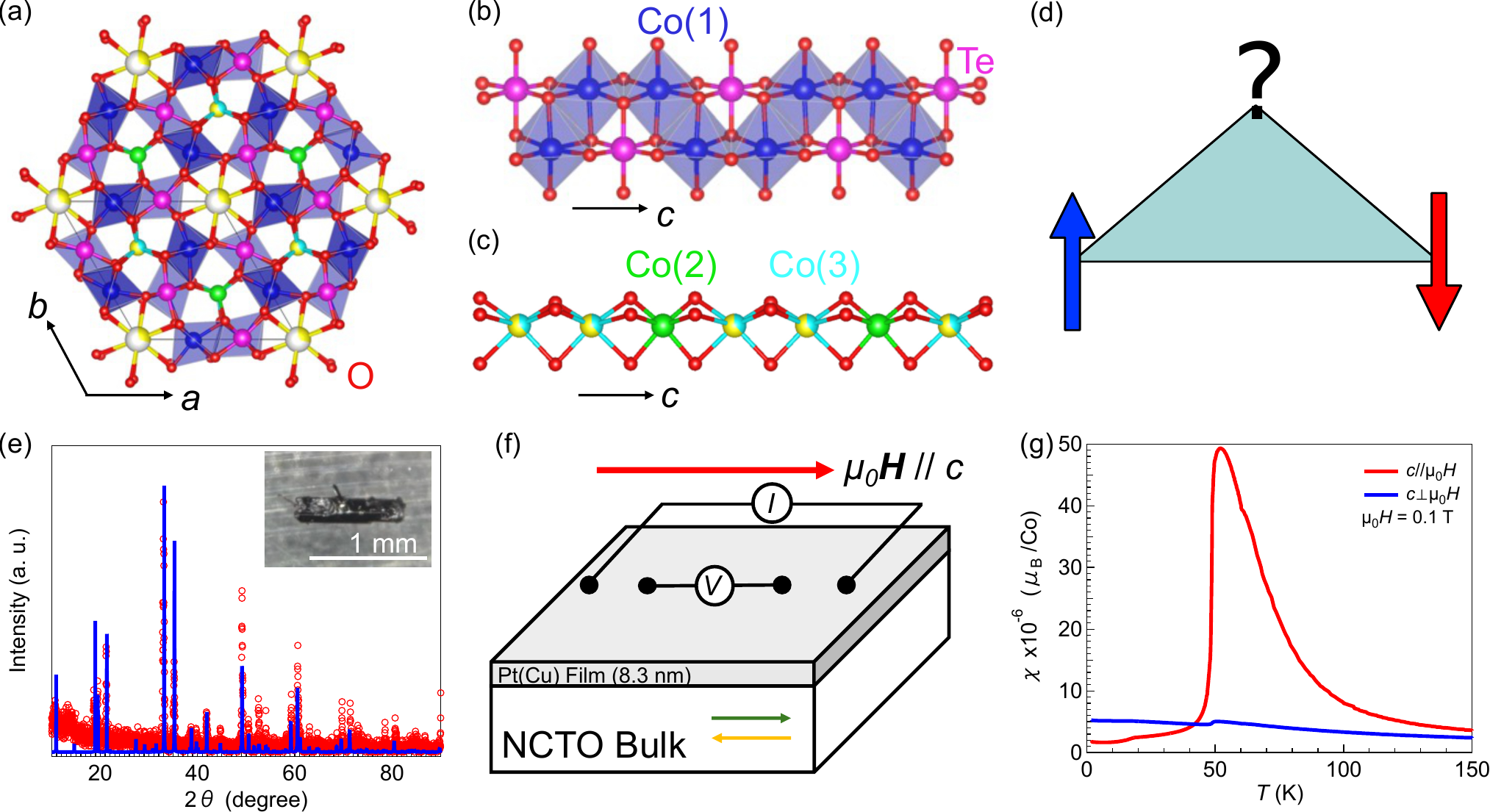}
    \caption{(a) Crystal structure of \ce{NCTO} projected in the $ab$ plane. Three \ce{Co} ion sublattices: \ce{Co(1)} (blue), \ce{Co(2)} (green), and \ce{Co(3)} (cyan). \ce{Co(2)} and \ce{Co(3)} are located at the center of the triangle formed by \ce{Co(1)}. (b) Zigzag chain of \ce{Co(1)} (blue) along the $c$-axis. (c) Alternate arrangement of \ce{Co(2)} (green) and \ce{Co(3)} (cyan) along the $c$-axis. \ce{Co(3)} site is partially replaced by \ce{Na} ions (yellow). (d) Magnetic structure of the \ce{Co(1)} triangle at low temperatures at zero field \cite{Zhao2022-hh}. (e) Powder X-ray diffraction pattern of \ce{NCTO} at room temperature. The red open circles represent the experimental data, and the blue lines represent the simulation results. The inset is a photograph of an \ce{NCTO} single crystal (850 $\times$ 340 $\times$ \SI{170}{\micro\meter^3}). (f) Illustration of the measurement setup for the longitudinal MR. The easy axis of the \ce{Co(1)} antiferromagnetic sublattice is oriented along the $c$-axis, to which electric current and magnetic field were applied. The green and orange arrows represent the sublattice magnetization.(g) Temperature dependence of dc magnetic susceptibility $\chi_{\parallel}$ and $\chi_{\perp}$, which were measured under the magnetic field applied parallel and perpendicular to the $c$-axis, respectively. The measurements were performed under field-cooling conditions at \SI{0.1}{\tesla}.}
    \label{fig:1}
\end{figure*}

The crystal structure of \ce{NCTO} projected on the $ab$-plane is shown in Fig. \ref{fig:1}(a). \ce{NCTO} possesses a hexagonal crystal structure belonging to the space group $P6_3/m$ \cite{Shan2014-gb,Zhao2022-hh,Saha2023-or}. The magnetic \ce{Co} ions are divalent and occupy three distinct crystallographic sites: \ce{Co(1)} (blue), \ce{Co(2)} (green), and \ce{Co(3)} (cyan). The \ce{Co(1)} sites with almost 80\% fraction dominate the magnetic properties \cite{Zhao2022-hh}. In the $ab$-plane, the \ce{Co(1)} ions form triangular network. The edge-sharing \ce{Co(1)O6} octahedra form zigzag chains along the $c$-axis [Fig. ~\ref{fig:1}(b)], which are suggested to exhibit ferromagnetic interactions within the chains. The \ce{Co(2)} and \ce{Co(3)} sites are located at the centers of the triangles of the \ce{Co(1)}, and the face-sharing \ce{Co(2)O6} and \ce{Co(3)O6} trigonal prisms alternate to form one-dimensional chains along the $c$-axis [Fig. ~\ref{fig:1}(c)]. 
\par

The Néel temperature ($T_{\text{N}}$) of \ce{NCTO} is approximately \SI{50}{\kelvin}, and the magnetic moments tend to align along the $c$-axis owing to the Ising anisotropy. This transition is governed by the dominant \ce{Co(1)} ions, and the anomaly just below $T_{\text{N}}$ due to the antiferromagnetic order of the \ce{Co(2)} and \ce{Co(3)} sublattices is rather weak in the magnetization curves \cite{Zhao2022-hh}. At $T_{\text{N}}$, the \ce{Co(1)} sublattice becomes a partially disordered antiferromagnetic (PDA) state~\cite{Zhao2022-hh}, where one of the \ce{Co(1)} spins in each \ce{Co(1)} triangle remains disordered, as illustrated in Fig. ~\ref{fig:1}(d). As temperature further decreases, the disordered moments become frozen, and a frozen PDA (FPDA) state emerges below approximately
\SI{20}{\kelvin} \cite{Zhao2022-hh}. When strong magnetic fields are applied from zero field along the $c$-axis, the magnetic moments of \ce{Co(1)} flip in the magnetic-field direction, and the induced ferromagnetic states are stabilized. The spin-flip transition is accompanied by a sudden change with clear hysteresis in the magnetic-field dependence of the magnetization \cite{Shan2014-gb,Zhao2022-hh,Saha2023-or}. The hysteresis during the spin-flip transition is notably large at very low temperatures \cite{Shan2014-gb,Zhao2022-hh,Saha2023-or}, which is a unique magnetic property of this material associated with the spin frozen state.
\par

In this study, we have investigated longitudinal MR, that is MR measured under a magnetic field applied parallel to the electric current, for metal/\ce{NCTO} bilayers. As metal layers, we selected \ce{Pt} with a strong SOI and \ce{Cu} with a negligible SOI to examine the MR contribution from the interface spin accumulation. In response to the change in the magnetic structure upon the spin-flip transition, we successfully observed MR anomalies during the spin-flip transition in both \ce{Pt/NCTO} and \ce{Cu/NCTO}. Even in the spin-frozen state at very low temperatures, the change in magnetic structures is detected sensitively as MR changes. The similar MR jumps observed for \ce{Pt/NCTO}, \ce{Pt/SiO_x/NCTO}, and \ce{Cu/NCTO} indicate that the origin of the MR anomalies is not interface spin accumulation but is attributed to the magnetostriction of \ce{NCTO}. A notable difference in the MR spectra between Pt and Cu was observed in the high-field regime, where the magnetic moments were fully aligned along the magnetic-field direction. The opposite signs of the slopes of the MR for \ce{Cu/NCTO} and \ce{Pt/NCTO} are prominent below $\sim$ \SI{50}{\kelvin}, suggesting a weak antilocalization effect due to the strong SOI of the Pt films.

\section{\label{sec:level1} Methods}
Single crystals of \ce{NCTO} were grown using a self-flux method, as described in a previous report \cite{Zhao2022-hh}. The starting materials, \ce{Na2CO3}, \ce{Co3O4}, and \ce{TeO2}, were mixed in a molar ratio of $3:2.33:6$. This mixture was thoroughly ground and mixed with ethanol, and then loaded into an alumina crucible. Subsequently, the powder was heated to \SI{1000}{\degreeCelsius} over \SI{10}{hours}, maintained at this temperature for \SI{2}{days}, and slowly cooled to \SI{600}{\degreeCelsius} at a cooling rate of \SI{2}{\degreeCelsius / h}. Finally, it was furnace-cooled to room temperature over \SI{7}{hours}. As shown in the inset to Fig.~\ref{fig:1}(e), the as-grown single crystals were black and elongated along the $c$-axis. The high phase purity of the obtained crystals was confirmed by powder X-ray diffraction [Fig. ~\ref{fig:1}(e)].

\ce{NCTO} is highly insulating at room temperature, and the resistivity is beyond the measurement limit of multimeters. The magnetization of a single-crystal piece was measured in the temperature range of 2 to \SI{300}{\kelvin} using a SQUID magnetometer (MPMS3, Quantum Design). The maximum magnetic field was \SI{7}{\tesla}. Magnetoresistance measurements were performed using a standard 4-terminal method in the temperature range of 2 to \SI{150}{\kelvin} using a Physical Property Measurement System (PPMS, Quantum Design) on crystal pieces of \ce{NCTO}, on which \ce{Pt} and \ce{Cu} thin films were deposited by magnetron sputtering methods. The thickness of the metal films was estimated to be 8.3 $\pm$ \SI{0.5}{\nano \meter} by X-ray reflectivity measurements performed on other films separately grown under the same sputtering conditions as those used in this study. Electrodes were formed on the metal films using silver paste. The measurement setup for the MR is shown in Fig. ~\ref{fig:1}(f). A longitudinal MR configuration was adopted, in which both the magnetic field and electric current were applied parallel to the $c$-axis. The magnetic field was swept between \SI{9}{\tesla} and \SI{-9}{\tesla}.

\section{\label{sec:level1} Results and Discussion}

\begin{figure*}[th]
    \centering
    \includegraphics[width=1\linewidth]{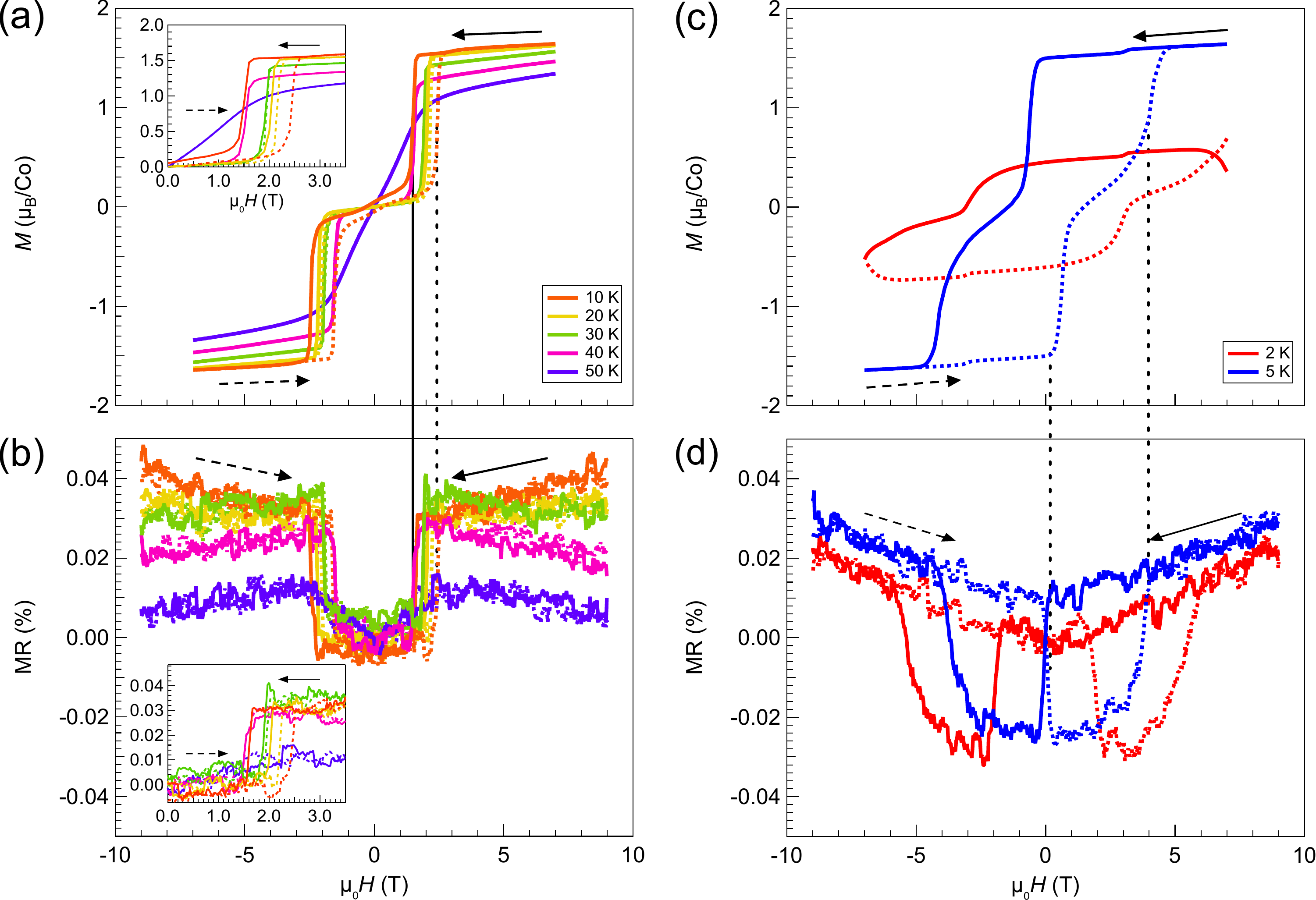}
    \caption{Magnetic field ($H$) dependence of (a),(c) magnetization ($M$) for an \ce{NCTO} single crystal and (b),(d) longitudinal MR spectra of a \ce{Pt/NCTO} bilayer at (a),(b) high temperatures ($\geq$ \SI{10}{\kelvin}) and (c),(d) low temperatures ($\leq$ \SI{5}{\kelvin}), respectively. The MR ratio is defined as MR(\%) $= \{ R(H) - R(0) \}/R(0) \times 100$, where $R(H)$ is the resistance at $H$. The insets of (a) and (b) show magnified views of magnetization and MR, respectively, in the range from $0$ to \SI{3.5}{\tesla}. In all panels, $H$ was applied parallel to the $c$-axis. Solid and dotted curves represent data obtained during field-decreasing (from \SI{9}{\tesla} to \SI{-9}{\tesla}) and field-increasing (from \SI{-9}{\tesla} to \SI{9}{\tesla}) scans, respectively. 
    The black solid and dashed lines drawn vertically between (a) and (b) and (c) and (d) are guides to compare the magnetization curves and MR. 
    }
    \label{fig:2}
\end{figure*}

Figure ~\ref{fig:1}(g) shows the temperature ($T$) dependence of the magnetic susceptibility when magnetic fields are applied parallel ($\chi_{\parallel}$) or perpendicular ($\chi_{\perp}$) to the $c$-axis. As the temperature decreases from \SI{150}{\kelvin}, $\chi_{\parallel}$ increases following the Curie-Weiss law, reaching a maximum at approximately \SI{50}{\kelvin}, after which it decreases rapidly. The overall temperature dependence is consistent with previous reports \cite{Shan2014-gb,Zhao2022-hh,Saha2023-or}. From the differential curve, the N\'eel temperature was estimated to be \SI{48.6}{\kelvin}. Upon further lowering the temperature, a small anomaly was observed near \SI{18}{\kelvin}, which can be ascribed to the transition to the frozen PDA (FPDA) state~\cite{Zhao2022-hh}. In contrast, the anomaly at the magnetic transitions in $\chi_{\perp}$ is much smaller than that in $\chi_{\parallel}$. This remarkable difference reflects a strong Ising anisotropy with an easy axis along the $c$-axis \cite{Zhao2022-hh}.
\par

Figures ~\ref{fig:2}(a) and (b) compare the magnetic-field ($H$) dependence of the magnetization ($M$) of \ce{NCTO} and that of the MR for the \ce{Pt/NCTO} heterostructure in the temperature range above \SI{10}{\kelvin}. Here, the magnetic field was applied parallel to the $c$-axis of the \ce{NCTO} crystal. As the temperature is lowered from \SI{50}{\kelvin}, steep changes in the magnetization at spin-flip transitions become noticeable. Below \SI{20}{\kelvin}, which is close to the transition temperature of the FPDA state, the sudden magnetization change at the spin-flip transition is accompanied by a hysteresis. This magnetization jump was first assigned as a spin-flop transition \cite{Shan2014-gb,Saha2023-or}, but was later interpreted as a spin-flip transition because of the strong Ising anisotropy \cite{Zhao2022-hh}.
The hysteresis during the spin-flip transition is likely associated with the spin-frozen state at very low temperatures, which may produce a pronounced history effect during the reorientation of the antiferromagnetic spins in the \ce{Co(1)} sublattice. In addition, the maximum magnetization increases gradually with decreasing temperature, reaching \SI{1.6}{\mu_B/\ce{Co}} at \SI{10}{\kelvin}. This increase in magnetization is due to the suppression of thermal fluctuations at low temperatures, which promotes spin alignment induced by an external magnetic field. The moment at \SI{7}{\tesla} is consistent with previous studies \cite{Zhao2022-hh, Shan2014-gb}, but it is only half of the spin-only value of \SI{3.0}{\mu_B/\ce{Co}} for the full moments of $d^7$ \ce{Co^2+} spins in the high-spin configuration. Splitting of the degenerate $d^7$ state due to the cooperative effect of the crystal field and SOI leads to a $J_{\text{eff}} = 1/2$ ground state, which explains the observed reduced magnetic moment \cite{Zhao2022-hh,Liu2020-nt,Kim2021-kj}.
\par

The MR spectra reflect the magnetization process and exhibit a clear magnetic-field dependence, as shown in Fig. \ref{fig:2}(b). The MR ratio, which is defined as $\{R(H)-R(H=0)\}/R(H=0) \times 100$\% where $R(H)$ is the resistance at $H$, increases sharply during the magnetization jump. The anomaly is more prominent at lower temperatures, and a clear hysteresis is observed below \SI{20}{\kelvin}, similar to the magnetization curves. The overall magnetic-field dependence of the MR resembles the isothermal magnetization curve. However, a key difference is that the high-field data exhibit nontrivial temperature variations. The slope of the MR ratio at \SI{50}{\kelvin} is negative, but it is positive at \SI{10}{\kelvin}. Such a sign change in the MR slope is usually not observed for \ce{Pt} single-layer films~\cite{Naumova2025-md}.  
\par

Below \SI{5}{\kelvin}, the magnetization shows a nontrivial magnetic-field dependence with one large hysteresis. It is known that isothermal magnetization curves are accompanied by hysteresis at spin-flip transitions \cite{Valenta2018-tb,Kim2022-vd}; however, the hysteresis in our case is remarkably wide. As shown in Fig.~\ref{fig:2}(c), two hysteresis loops in positive and negative magnetic fields merge, and only one large hysteresis is observed in the magnetization curve. As reported with surprise \cite{Shan2014-gb,Zhao2022-hh,Saha2023-or}, the field-induced ferromagnetic state remains stable down to zero field, which has a remanent magnetization with a magnitude of about \SI{1.5}{\mu_B/\ce{Co}}. Although \ce{NCTO} is antiferromagnetic, magnetization curves similar to those of ferromagnets are observed. At \SI{2}{\kelvin}, the field where the magnetization saturates is even larger than the maximum magnetic field of the SQUID magnetometer ($\pm$ \SI{7}{\tesla}), and the characteristic of a minor loop is observed. This minor-loop behavior is more noticeable in our samples than in previous reports \cite{Shan2014-gb,Zhao2022-hh,Saha2023-or}, possibly due to the different purity of the grown samples. Several mechanisms, such as a metastable ground state and kinetic arrest of a first-order antiferromagnetic-ferromagnetic transition arising from frozen magnetic structure, have been discussed as the origin of this significant irreversible magnetization \cite{Zhao2022-hh}, but the origin has not yet been established.
\par

Irrespective of the microscopic origin, the spin freezing crucial at very low temperatures is likely to be relevant to the large hysteresis. Even in such a spin-frozen state with complicated magnetic interactions, the MR sensitively detects the change in the magnetic structure, as shown in Fig. ~\ref{fig:2}(d). We observed MR spectra with a peculiar shape having two hysteresis loops, which is similar to tunnel magnetoresistance in which large MR contrast was observed between parallel and anti-parallel arrangements of two ferromagnetic layers \cite{Miyazaki1995-mz,Moodera1995-rf}. When the magnetic field was swept from positive to negative (negative to positive), a dip was observed in the negative (positive) magnetic-field region.

Although the magnetic-field dependence of MR does not follow that of magnetization in this very low-temperature regime, 
the trend is qualitatively understood in the same way as the MR above \SI{10}{\kelvin}; namely, the resistivity is higher in the induced ferromagnetic state than in the antiferromagnetic state, resulting in the jump of the MR. For example, at \SI{5}{\kelvin}, a positive MR with an almost linear slope was observed in the induced ferromagnetic state, and negative dips were observed only during the intermediate state between the positively and negatively polarized states. In the intermediate state, the \ce{Co(1)} moments are not perfectly antiferromagnetic because of the spin freezing, but the MR sensitively detects the magnetic-structure change from the induced ferromagnetic state.

\begin{figure*}[t]
    \centering
    \includegraphics[width=1\linewidth]{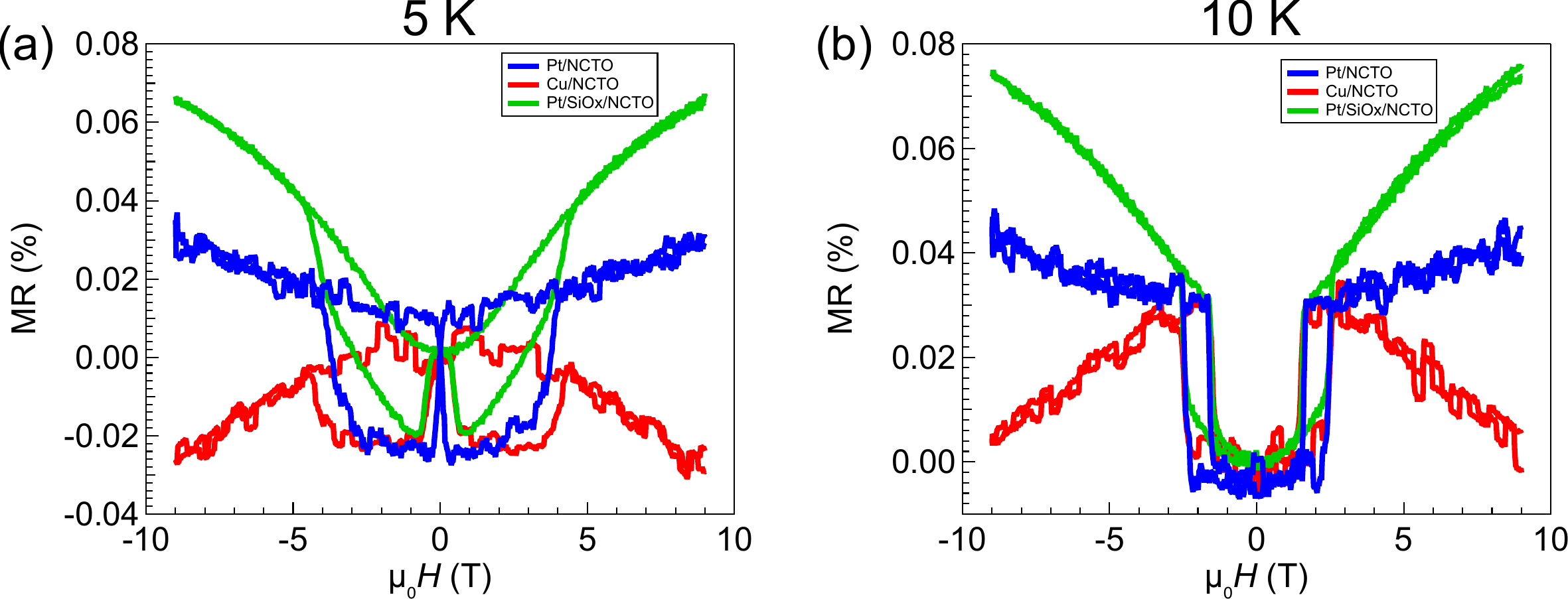}
    \caption{Magnetic field ($H$) dependence of the MR ratio of \ce{Pt/NCTO} (blue), \ce{Cu/NCTO} (red), and \ce{Pt/SiO_x/NCTO} (green) at (a) \SI{5} {\kelvin} and (b) \SI{10} {\kelvin}. The MR ratio is determined by MR(\%) $= \{ R(H) - R(0) \}/R(0) \times 100$, where $R (H)$ is the resistance at $H$. The data for \ce{Pt/NCTO} are the same as those already shown in Figs.~\ref{fig:2} (b) and (d).}
    \label{fig:3}
\end{figure*}

To check whether the origin of the abrupt changes in the MR at the spin-flip transition is associated with interface spin accumulation by the spin Hall effect, we performed the same MR measurements on a \ce{Cu/NCTO} heterostructure, where \ce{Cu} possesses a negligible SOI. Figures ~\ref{fig:3}(a) and (b) show a comparison of the MR spectra for \ce{Pt/NCTO} and \ce{Cu/NCTO} at \SI{5}{\kelvin} and \SI{10}{\kelvin}, respectively. We observed negative dips with similar magnitudes for both \ce{Cu/NCTO} and \ce{Pt/NCTO} at both temperatures. This suggests that the MR jump is not due to spin-current effects such as the SMR, because the spin current generated via the spin Hall effect is negligibly small in \ce{Cu}.

A plausible mechanism of the sudden resistance changes in both \ce{Pt/NCTO} and \ce{Cu/NCTO} is the magnetostriction. To confirm this scenario, we further measured the MR of a \ce{Pt/SiO_x/NCTO} heterostructure, where the insulating \ce{SiO_x} layer with a thickness of \SI{20}{\nano \meter} blocks spin current transmission. As shown in Fig. ~\ref{fig:3}, negative dips similar to those observed for \ce{Pt/NCTO} and \ce{Cu/NCTO} were observed at both \SI{5}{\kelvin} and \SI{10}{\kelvin}. Similar results obtained for samples in which the interface exchange interaction is eliminated support the magnetostriction scenario.

The dimensions of the metal/\ce{NCTO} heterostructure can slightly change at the spin-flip transition via magnetostriction, which affects the resistance of the metal films.  
We estimate the magnetostriction coefficient $\lambda$ from the MR data following the procedure in a previous study on \ce{Pt}/\ce{Cr2O3} (\SI{5}{\nano \metre}) bilayers \cite{Liao2025-hu}. Assuming that the observed MR jump originates purely from the geometric change induced by the magnetostriction of \ce{NCTO}, the resistance change ratio of the adjacent metal film is approximated by $2\Delta L/L$ \cite{Liao2025-hu}, where $L$ is the sample length and $\Delta L$ is the variation in the sample length due to the magnetostriction. Since $\lambda$ is defined as $\Delta L/L$, $\lambda$ can be estimated from the change in the resistance. We obtain $\lambda = 1.4 \times 10^{-4}$ from the \SI{10}{\kelvin} data.
This value is more than an order of magnitude larger than the reported values at the spin-flop transition for \ce{NiO} \cite{Alberts1961-jh} and \ce{Cr2O3} \cite{Liao2025-hu}, but is an order of magnitude smaller than that reported for \ce{MnTe} \cite{Baral2023-gg}. In addition, the MR change observed for \ce{Pt/Cr2O3} at the spin-flop transition was also positive, similar to our case \cite{Liao2025-hu}. Hence, the magnetostriction scenario is feasible.
\par

Although the MR jumps at spin-flip transitions are similar for \ce{Pt/NCTO} and \ce{Cu/NCTO}, the MR spectra in the high magnetic-field regime are interestingly different between these two samples; the slope for \ce{Pt/NCTO} is positive, while it is negative for \ce{Cu/NCTO} at 5 and \SI{10}{\kelvin} (Fig. ~\ref{fig:3}). In this high-field region, the moments of \ce{NCTO} are fully polarized along the magnetic-field direction. In contrast to the low-field MR anomaly at the spin-flip transition, the different MR spectra in this high-field region are associated not with the magnetic-structure change of \ce{NCTO}, but with the material of the metal layers.
\par

\begin{figure}[t]
    \centering
    \includegraphics[width=1\linewidth]{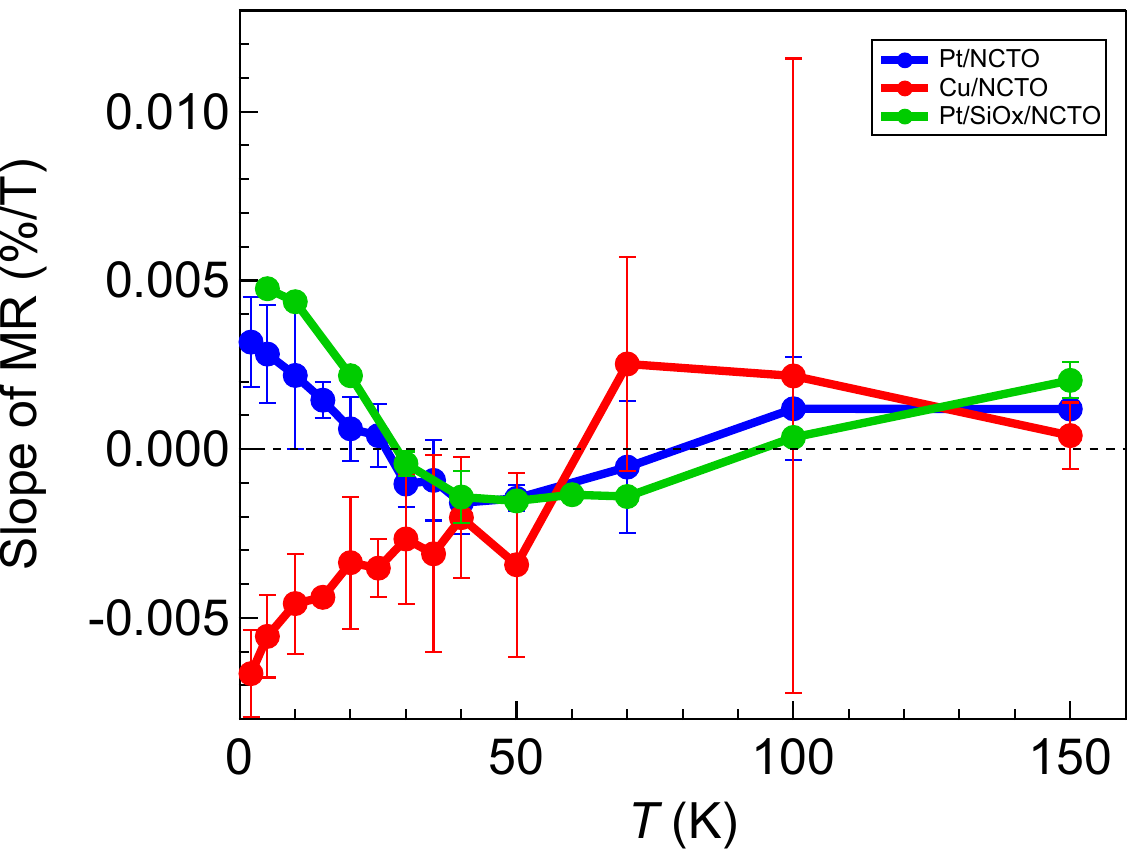}
    \caption{Temperature ($T$) dependence of the slope of MR spectra at high fields ($\mu_0|H| \ge$ \SI{7}{\tesla}) for \ce{Pt/NCTO} (blue), \ce{Cu/NCTO} (red), and \ce{Pt/SiO_x/NCTO} (green). The plotted data are the averaged results obtained from four field sweeps (9 $\rightarrow$ \SI{7}{\tesla}, $-7 \rightarrow$ \SI{-9}{\tesla}, $-9 \rightarrow $ \SI{-7}{\tesla}, and 7 $\rightarrow$ \SI{9}{\tesla}), where the sign of the negative field data is inverted. Error bars indicate the standard deviation in the linear fit.}
    \label{fig:4}
\end{figure}

The temperature dependence of the MR slope in the high-field region is shown in Fig. ~\ref{fig:4}. Here, the slope was estimated by the average of the linear fits to the MR data from \SI{9}{\tesla} to \SI{7}{\tesla}, from \SI{-7}{\tesla} to \SI{-9}{\tesla}, from \SI{-9}{\tesla} to \SI{-7}{\tesla}, and from \SI{7}{\tesla} to \SI{9}{\tesla}. 
The slope for \ce{Cu/NCTO} is positive at \SI{150}{\kelvin} but gradually decreases and becomes negative below about \SI{60}{\kelvin}, although the high-temperature data are accompanied by large error bars because of the very small MR. For \ce{Pt/NCTO} and \ce{Pt/SiO_x/NCTO}, a similar trend was observed down to approximately \SI{50}{\kelvin}, but it starts to increase below that temperature. This indicates that the positive MR contribution becomes dominant in \ce{Pt/NCTO} and \ce{Pt/SiO_x/NCTO} below \SI{50}{\kelvin}. The difference in the MR slope between \ce{Pt/NCTO} and \ce{Cu/NCTO} becomes prominent below $T_{\text{N}}$(=\SI{48.6}{\kelvin}), suggesting that the magnetic order of \ce{NCTO} is related to the different MR slopes between \ce{Pt/NCTO} and \ce{Cu/NCTO}.
\par

In discussing the different temperature dependence of the MR between \ce{Pt/NCTO} and \ce{Cu/NCTO}, one key characteristic of \ce{Pt} is the Stoner-enhanced Pauli paramagnetic nature, which is absent in \ce{Cu}. Since the band structure of \ce{Pt} is close to ferromagnetic, the interface between \ce{Pt} and \ce{NCTO} could be magnetized owing to the magnetic proximity effect as reported for \ce{Pt/Y_3Fe_5O_{12}} \cite{Huang2012-jm}, resulting in MR in the \ce{Pt} layer. If the induced magnetization of \ce{Pt} gives rise to anisotropic magnetoresistance, a positive MR can be expected when an electric current is applied parallel to the magnetization \cite{Huang2012-jm}. However, for the \ce{SiO_x}-inserted sample in which the magnetic proximity effect should vanish, a positive MR with a magnitude similar to that of \ce{Pt/NCTO} was observed (Figs.~\ref{fig:3} and \ref{fig:4}). Hence, the MR originating from the induced magnetization of \ce{Pt} via the magnetic proximity effect is unlikely.  
\par

A strong SOI is another distinctive property of \ce{Pt} compared to \ce{Cu}. The interface spin accumulation due to the spin Hall effect in \ce{Pt} may explain the positive MR at high fields, since it should become prominent below $T_{\text{N}}$ owing to the interaction between the spin accumulation and the induced ferromagnetic moments of \ce{NCTO}. In heavy metal/magnetic insulator bilayers, a typical MR effect caused by the interface spin accumulation is the SMR. In our configuration of the MR measurements, however, the SMR is not active; the SMR occurs in the presence of the N\'eel vector component perpendicular to the electric current, while the N\'eel vector is parallel to the electric current in our case owing to the strong Ising anisotropy. Furthermore, a smilar positive MR was observed also for \ce{Pt/SiO_x/NCTO}, where the insulating \ce{SiO_x} layer blocks the transmission of spin current across the NCTO interface.

Another possibility is the Hanle MR. The Hanle MR is expected because the applied magnetic field points perpendicular to the spin accumulation. Dephasing of spin accumulation in magnetic fields suppresses the backflow of spin currents, leading to a positive MR \cite{Velez2016-er}. Notably, the Hanle MR can occur for metal films on nonmagnetic substrates, and its magnitude sensitively depends on the film roughness and average grain size \cite{Velez2016-er}. In this context, the similar high-field MR spectra of \ce{Pt/NCTO} and \ce{Pt/SiO_x/NCTO} are consistent with the Hanle MR. However, the resistivity modulation of the metal film is caused by magnetic fields \cite{PhysRevLett.99.126601} instead of the magnetization of the underlying layer in the Hanle MR. Thus, it is difficult to explain why the positive MR is only pronounced below \SI{50}{\kelvin} ($\approx T_ \text{N}$).

Rather, the positive MR plausibly originates from the weak anti-localization (WAL)~\cite{hnl1980,bergman1982}. Since our metal films were prepared by magnetron sputtering at room temperature, disorders inevitably introduced during the deposition can result in WAL. Although the WAL is not related to the magnetism of \ce{NCTO} and thus the onset temperature of WAL is not linked to $ T_\text{N}$ of \ce{NCTO}, a previous paper \cite{Velez2016-er} for \ce{Pt/Y_3Fe_5O_{12}} reports that the WAL appears below \SI{50}{\kelvin}, close to $T_\text{N}$ of \ce{NCTO}. The magnitude of the WAL at \SI{10}{\kelvin} for \ce{Pt/Y_3Fe_5O_{12}} is on the order of \SI{0.01}{\%} at \SI{9}{\tesla} \cite{Velez2016-er}, similar to that of the positive MR in our \ce{Pt/NCTO} (Fig. ~\ref{fig:3}). From the WAL fit of the data at \SI{5}{\kelvin}, the phase coherence length of \ce{Pt} is estimated to be approximately \SI{55}{\nano\meter} for \ce{Pt/NCTO} and \SI{50}{\nano\meter} for \ce{Pt/SiO_x/NCTO}, consistent with the values reported for Pt films \cite{Jana2023-sj}. The difference in the positive MR magnitude between \ce{Pt/NCTO} and \ce{Pt/SiO_x/NCTO} at low temperatures may be ascribed to different \ce{Pt}-film qualities.

\par

Finally, we note that the negative MR observed in \ce{Cu/NCTO} below $T_{\text{N}}$ is also nontrivial. Usually, the MR in non-magnetic metals is positive owing to the Lorentz force effect. Weak localization \cite{Van-Haesendonck1981-sf,Rosenbaum1983-ct} is a possible mechanism for the negative MR in the \ce{Cu/NCTO} at high fields, but the negative MR is also observed in \ce{Pt/NCTO} and \ce{Pt/SiO_x/NCTO} with a large SOI in the intermediate temperature region (from 30 to \SI{70}{\kelvin}). Thus, this scenario is unlikely at least in the intermediate temperature region. Although a naturally oxidized \ce{Cu} layer may exhibit orbital Hall effect \cite{PhysRevLett.121.086602}, the negative MR cannot be explained by the orbital Hanle MR, whose sign is positive \cite{PhysRevLett.131.156703} similar to the spin Hanle MR.

The negative MR may be ascribed to scattering due to interface roughness \cite{Wang2021-mt,PhysRevLett.87.126805} because as-grown single crystals of \ce{NCTO} were used. Atomic force microscopy observations revealed that the surface root-mean-square roughness of the as-grown \ce{NCTO} crystals was at most \SI{4}{nm}. Here, the scanned area was \SI{5}{\micro\metre} $\times$ \SI{5}{\micro\metre} and similar roughness values were obtained in several scanned areas. Since the \ce{Pt} and \ce{Cu} films were deposited on the as-grown \ce{NCTO} crystals and its thickness is \SI{8.3}{\nano\metre}, this small roughness may be critical for electron scattering. In addition, the roughness of the top surface of the metal films may also affect the electron scattering. Hence, roughness-induced scattering at the interfaces and the film surfaces could contribute to the negative MR, especially at low temperatures where electron–phonon scattering is suppressed and the mean free path becomes long. We note that the resistivities of \ce{Pt} and \ce{Cu} films at room temperature are \SI{12.1}{\micro\ohm\centi\meter} and \SI{22.9}{\micro\ohm\centi\meter}, respectively. The resistivity of the \ce{Pt} film is close to literature values, whereas that of the \ce{Cu} film is significantly higher than the bulk value ($\sim$ \SI{1}{\micro\ohm\centi\meter} \cite{Kim2019-tw}). This may be consistent with the strong roughness scattering as the origin of the negative MR in \ce{Cu/NCTO}. In \ce{Pt/NCTO}, the negative MR due to the roughness effect is significant in the intermediate temperature region, but the SOI effect, probably WAL, becomes dominant below approximately \SI{50}{\kelvin}.

\section{\label{sec:level1} Conclusions}

In this study, we fabricated \ce{Pt} and \ce{Cu} thin films on \ce{NCTO} single crystals and investigated the magnetic-field dependence of the MR in a geometry where both the electric current and magnetic field were applied parallel to the $c$-axis of the \ce{NCTO} crystal. Clear anomalies reflecting the spin-flip transition were observed in the MR spectra below the N\'eel temperature. The magnetization process of the \ce{Co(1)} sublattice was sensitively detected by MR, even at low temperatures where the freezing of \ce{Co} spins is prominent. Since MR anomalies with similar magnitudes were observed at the spin-flip transition for \ce{Pt/NCTO}, \ce{Pt/SiO_x/NCTO}, and \ce{Cu/NCTO}, the origin is attributed to deformation-induced strain effects caused by the magnetostriction of \ce{NCTO}. The important difference between the MR spectra of \ce{Pt/NCTO} and \ce{Cu/NCTO} is the opposite sign of the slope of the MR in the induced ferromagnetic state of \ce{NCTO} at low temperatures. Based on the similar MR spectra observed for the \ce{SiO_x}-inserted sample, we conclude that the high-field MR in \ce{Pt/NCTO} is dominated by weak anti-localization at low temperatures, whereas interface/surface roughness plays an important role in the negative MR in the intermediate temperature region. The successful electrical detection of the magnetic properties of an underlying Ising antiferromagnet could contribute to the further development of antiferromagnetic spintronics.

\begin{acknowledgments}
The authers are grateful to Dr. Masataka Kawano for fruitful discussion. A part of this work was carried out by the joint research of the Cryogenic Research Center, the University of Tokyo. This work was supported by the Murata Science and Education Foundation (No. M25AN137), by JST FOREST Program, Grant No. JPMJFR203H, and by JSPS KAKENHI, Grants No. JP26K01375, No. JP26H00629, No. JP24K00566, and No. JP25H00611.

\end{acknowledgments}





\bibliographystyle{style}
\bibliography{ref}

\end{document}